\documentclass[12pt, letterpaper]{article}
\usepackage[top=1.2in, bottom=1.5in, left=1.2in, right=1.2in]{geometry}
\usepackage[utf8]{inputenc}
\usepackage{booktabs}
\usepackage{hyperref}
\usepackage{amsmath}
\usepackage{amssymb}
\usepackage{natbib}
\usepackage{graphicx}

\title{LSST Observations of {\it WFIRST} Deep Fields}

\author{The WFIRST Deep Field Working Group,\\
WFIRST Science Investigation Team Members,\\
and Community Members:\\
R.\ J.\ Foley, A.\ M.\ Koekemoer, D.\ N.\ Spergel,\\
F.\ B.\ Bianco,
P.\ Capak,
L.\ Dai,
O.\ Dor{\'e},
G.\ G.\ Fazio,
H.\ Ferguson,\\
A.\ V.\ Filippenko,
B.\ Frye,
L.\ Galbany,
E.\ Gawiser,
C.\ Gronwall,\\
N.\ P.\ Hathi,
C.\ Hirata,
R.\ Hounsell,
S.\ W.\ Jha,
A.\ G.\ Kim,\\
P.\ L.\ Kelly,
J.\ W.\ Kruk,
S.\ Malhotra,
K.\ S.\ Mandel,
R.\ Margutti,\\
D.\ Marrone,
K.\ B.\ W.\ McQuinn,
P.\ Melchior,
L.\ Moustakas,\\
J.\ A.\ Newman,
J.\ E.\ G.\ Peek,
S.\ Perlmutter,
J.\ D.\ Rhodes,\\
B.\ Robertson,
D.\ Rubin,
D.\ Scolnic,
R.\ Somerville,
R.\ Street,\\
Y.\ Wang,
D.\ J.\ Whalen,
R.\ A.\ Windhorst,
E.\ J.\ Wollack}

\date{November 30, 2018}

\begin{document}

\maketitle

\begin{abstract}
The Wide-Field Infrared Survey Telescope ({\it WFIRST}) is expected to launch in the mid-2020s.  With its wide-field near-infrared (NIR) camera, it will survey the sky to unprecedented detail.  As part of normal operations and as the result of multiple expected dedicated surveys, {\it WFIRST} will produce several relatively wide-field (tens of square degrees) deep (limiting magnitude of 28 or fainter) fields.  In particular, a planned supernova survey is expected to image 3 deep fields in the LSST footprint roughly every 5 days over 2 years.  Stacking all data, this survey will produce, over all {\it WFIRST} supernova fields in the LSST footprint, $\sim$12--25 deg$^{2}$ and $\sim$5--15 deg$^{2}$ regions to depths of $\sim$28~mag and $\sim$29~mag, respectively.  We suggest LSST undertake mini-surveys that will match the {\it WFIRST} cadence and simultaneously observe the supernova survey fields during the 2-year {\it WFIRST} supernova survey, achieving a stacked depth similar to that of the {\it WFIRST} data.  We also suggest additional observations of these same regions throughout the LSST survey to get deep images earlier, have long-term monitoring in the fields, and produce deeper images overall.  These fields will provide a legacy for cosmology, extragalactic, and transient/variable science.
\end{abstract}

\newpage
\section{White Paper Information}
Corresponding authors: The full list of authors is listed above.  Inquiries can be directed to R.~J.~Foley (foley@ucsc.edu), A.~M.~Koekemoer (koekemoer@stsci.edu), and D.~N.~Spergel (dns@astro.princeton.edu).
 %In addition, please provide the following categorization for your white paper:
\begin{enumerate} 
\item {\bf Science Category:} 
Our proposed observations would directly address the ``constraining dark energy and dark matter'' and ``exploring the transient optical sky'' science goals outlined in the LSST SRD.  The observations would also significantly improve extragalactic science and improve calibration efforts for both LSST and {\it WFIRST}.
\item {\bf Survey Type Category:} Our proposal falls under the category of mini survey since the observations do not need to be performed for the entire 10-year LSST survey.  However, during the 2-year period, the observations are similar to the deep-drilling field observations. 
\item {\bf Observing Strategy Category:} We request an integrated strategy to observe the specific {\it WFIRST} supernova deep fields during the 2-year {\it WFIRST} supernova survey.  Ideally, the cadence and timing of the observations would be matched to that of {\it WFIRST}.  We would prefer observations in all LSST filters to deep-drilling field depths. A combined {\it WFIRST}--LSST supernova survey strategy also satisfies the science requirements for a wide variety of other scientific applications that can be studied with deep extragalactic fields.
\end{enumerate}

\clearpage

\section{Scientific Motivation}
%\begin{footnotesize}
%{\it Describe the scientific justification for this white paper in the context
%of your field, as well as the importance to the general program of astronomy, 
%including the relevance over the next decade. 
%Describe other relevant data, and justify why LSST is the %best facility for these observations.
%(Limit: 2 pages + 1 page for figures.)}
%\end{footnotesize}

{\it WFIRST} is the top space-based priority from the 2010 Decadal survey and will be the flagship NASA mission following the James Webb Space Telescope ({\it JWST}).  {\it WFIRST} has a 2.4-m mirror and a highly capable instrument suite \citep[for a detailed description, see][]{Spergel15}.  For the purposes of this white paper, we focus on the Wide-Field Instrument (WFI), a 0.28-deg$^{2}$ imager with a complement of several wide filters ($RZY\!J\!H\!F$, and one ultra-wide) and a grism.  It will have an angular resolution comparable to the {\it Hubble Space Telescope}, but with a field of view that is $\sim$100 times larger.

Following the recommendations of the 2010 Decadal survey, the {\it WFIRST} Project is preparing for several surveys to address science topics related to exoplanets and dark energy.  In particular, we expect a relatively shallow High-Latitude Survey (HLS) covering $\sim$2000~deg$^{2}$ overlapping the LSST footprint and a Supernova (SN) survey that will repeatedly visit $\sim$40~deg$^{2}$.

The {\it WFIRST} SN survey \citep[see][]{Hounsell18} is expected to have a cadence of $\sim$5 days over a period of 2 years (the middle 2 years of the 5-year mission).  It is expected to observe 20--40~deg$^{2}$ to $\sim$22~mag in single epochs (the ``shallow'' fields) and 5--15~deg$^{2}$ to $\sim$25--26~mag in single epochs (the ``deep'' fields).  The shallow and deep fields will likely be observed in $RZY\!J$ (and possibly $H$) and $Y\!J\!H\!F$ (and possibly $Z$), respectively.  Stacking all exposures over the 2-year survey, we expect the shallow and deep fields to reach depths of $\sim$28 and 29~mag in each filter, respectively.

With these data, we will discover $\sim$20,000 Type Ia supernovae (SNe~Ia).  With the SN light-curve data, we will constrain the expansion history of the Universe to $z \approx 3$.  We will detect $\sim$8000 SNe~Ia at $z < 1$, for which LSST can obtain complementary $ugrizy$ light curves, resulting in data from 0.3--2~$\mu$m.  The combined optical/NIR light curves will (1) improve the distance precision of these SNe beyond what either telescope could do alone, (2) provide strong systematic tests by tracking the rest-frame optical from $z = 0$ to $z = 3$, (3) increase the wavelength coverage, constraining dust properties, and (4) provide end-to-end testing of transient object detection and characterization at the faintest magnitudes LSST probes.  Using the {\it WFIRST} prism, we will obtain redshifts and classifications for a subset of these SNe.

Additionally, these fields will be used to calibrate photometric redshifts (photo-$z$'s), critical for several science cases for LSST and {\it WFIRST}.  To maximize the utility of these data, having similar depths in both LSST and {\it WFIRST} filters is required \citep{Hemmati18}.  The additional {\it WFIRST} photometry and high-resolution images will break photo-$z$ degeneracies and improve the precision for all galaxies.  With the high-resolution imaging, one can better deblend galaxies and construct empirical models how blends affect photo-$z$’s.  Finally, the deeper imaging will result in significantly higher signal-to-noise measurements for galaxies of a given luminosity than the combination of HLS and the WFD surveys.

Deep spectroscopic observations are needed to calibrate the {\it WFIRST} HLS grism survey. To obtain 1\% redshift purity at the {\it WFIRST} HLS flux limit, 22~deg$^{2}$ of deep grism spectroscopy are planned with at least 10 observation sets.  Each set, which is designed to duplicate the main HLS grism survey observations, is obtained at 4 different roll angles, resulting in a total of $\ge$40 separate spectral observations (each at a different roll angle) of these fields.  We expect these calibration fields to coincide with the SN fields that overlap with the HLS.  We expect these observations will yield a 99.9\% completeness with negligible confusion.  The HLS grism survey calibration observations will provide spectroscopic redshifts for SN host galaxies, and enable the calibration of photometric redshifts for weak lensing observations for both {\it WFIRST} and LSST.

The {\it WFIRST} deep fields will establish a unique and important legacy for extragalactic science, covering a range of scientific questions from reionization through the peak epoch of star formation.  Coordinated {\it WFIRST}--LSST deep fields would provide exquisite imaging in $\sim$10 filters to sufficient depth and area to tackle an enormous range of scientific questions from reionization through the peak epoch of star formation. The statistical returns for the galaxy populations yielded by such a survey are transformational.  In the deep fields, where deep LSST imaging will be critical for photo-z vetoes, we will discover $\sim${}$10^{5}$ $z\approx8$ galaxies. These objects will be selected over an area and depth that will uniquely complement redshifted 21-cm observations of the IGM neutral fraction, allowing for cross-correlations between galaxies and the IGM ionization state that will crystallize our understanding of how reionization unfolded.  We will obtain rest-frame optical observations for $z \lesssim 3$, providing precise stellar masses that can be compared to their rest-frame UV derived star-formation rates.  We will simultaneously constrain the abundance of the most massive galaxies at $z > 4$, which will provide constraints on the efficiency of star formation at these epochs, and will allow us to quantify the demographics of the first galaxies to cease their star formation, yielding insights into the physical mechanisms responsible for quenching.

Combining the LSST photometry with {\it WFIRST} grism spectra  will enable H$\alpha$+[O\,{\sc III}] observations for $\sim${}$10^{6}$ galaxies at $z \approx 2$, enabling a cross correlation between star-formation rate, rest-frame UV emission, high-resolution rest-frame optical morphology, and cosmic environment/spatial clustering. The abundance of relatively low-luminosity active galactic nuclei and their connection to their host galaxies can be measured for samples of unprecedented size at $z > 4$. Further combining with the weak-lensing-based estimates of dark matter halo mass, these data will provide definitive measures of the relationship between galaxy and dark matter halo properties and the connection between star formation, stellar mass, and halo growth, leading to invaluable constraints on theoretical models of galaxy formation.  The {\it WFIRST}--LSST collaborative survey efforts therefore have the potential to rewrite the origin story of modern galaxies.

%\vspace{.6in}
\newpage

\begin{figure}[h!]
 \begin{minipage}[c]{1.0\linewidth}
  \centering
  \begin{center}
  \includegraphics[width=0.95\linewidth, angle=0]{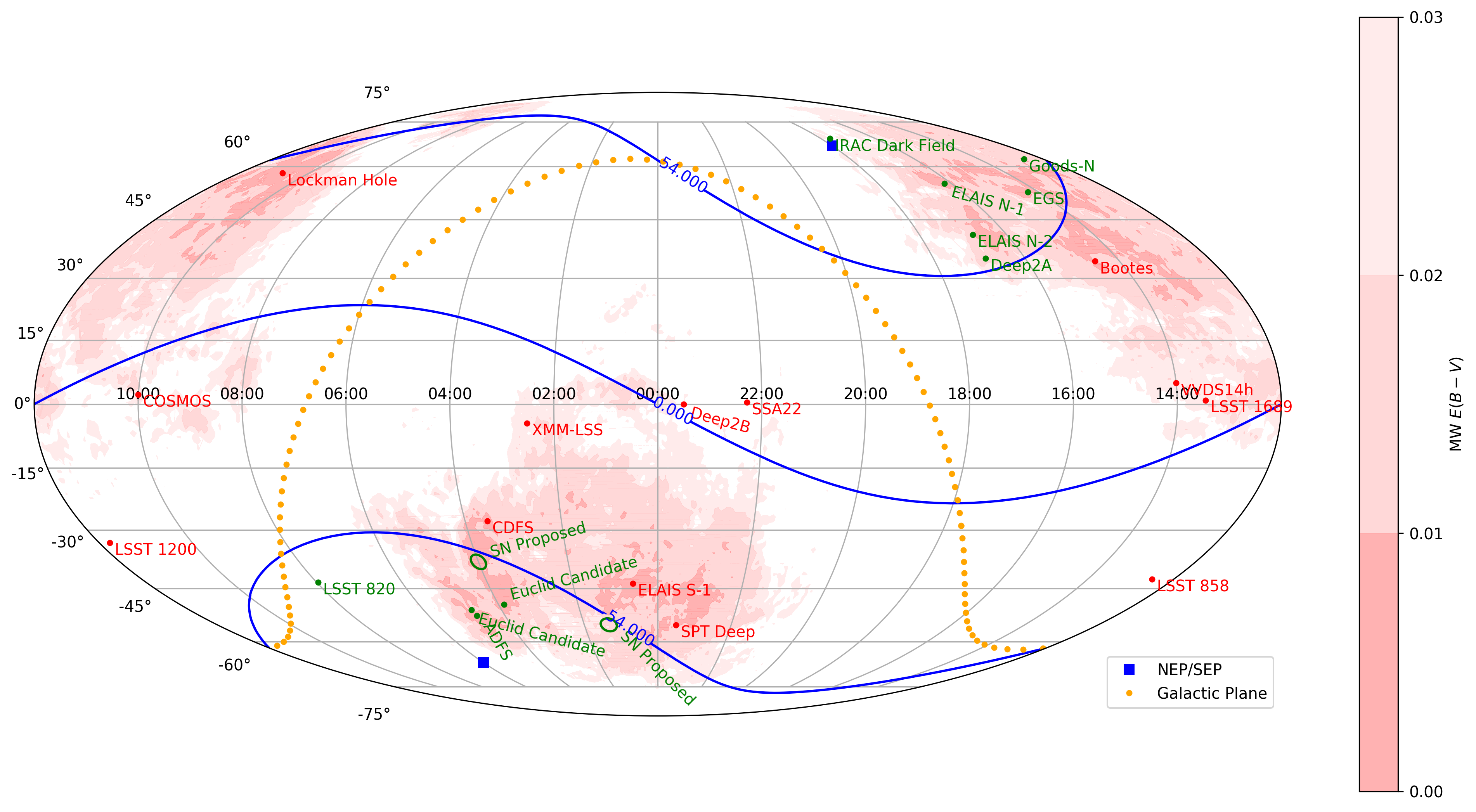}
  \vspace{-5pt}
  \caption{Equatorial map of the sky indicating potential SN fields for {\it WFIRST} and LSST.  The red shading displays the Milky Way reddening (as indicated by the color bar on the right) with lower reddening values being darker.  The Galactic plane is represented by the gold dots.  The ecliptic plane is indicated by the central blue line (labeled 0.000) and the ecliptic poles are marked with blue squares.  Ecliptic latitudes of $|54^{\circ}|$, corresponding to the edge of the {\it WFIRST} continuous viewing zone (CVZ) are also displayed as blue lines.  Several extragalactic fields are marked with those in green being in the {\it WFIRST} CVZ.  The three proposed {\it WFIRST} SN deep fields overlapping the LSST footprint are the AKARI Deep Field South (labeled ``ADFS''), which will also be observed by Euclid, and those labeled ``SN Proposed.''}\label{f:fields}
  \end{center}
  \end{minipage}
\end{figure}
\vspace{-2em}

\newpage
\section{Technical Description}
%\begin{footnotesize}
%{\it Describe your survey strategy modifications or proposed observations. Please comment on each observing constraint
%below, including the technical motivation behind any constraints. Where relevant, indicate
%if the constraint applies to all requested observations or a specific subset. Please note which 
%constraints are not relevant or important for your science goals.}
%\end{footnotesize}

\subsection{High-level description}
%\begin{footnotesize}
%{\it Describe or illustrate your ideal sequence of observations.}
%\end{footnotesize}

We request ``simultaneous'' observations of all {\it WFIRST} SN fields in the LSST footprint.  The {\it WFIRST} observing sequence is currently designed to take 30 hours every 5 days, making truly simultaneous observations impossible for all fields.  However, it is likely that the different fields can be somewhat staggered, allowing planned LSST observations to be within 12 hours of {\it WFIRST} observations.

We request a sequence of ``deep-drilling''-type depth exposures for each visit, reaching $u \approx 23.5$, $g \approx 25.3$, $r \approx 25.6$, $i \approx 25.4$, $z \approx 24.9$, and $y \approx 24.0$~mag (depending on the lunar phase).  All filters could be observed in a single visit, but observing all filters over a few nights would be acceptable as long as at least two filters (one color) is observed each night of observations.  The stacked depth of these fields would be slightly shallower than the final stacked depths of deep drilling fields ($ugrizy = 26.8$, 28.4, 28.5, 28.3, 28.0, 26.2~mag, respectively).

Since {\it WFIRST} will obtain deep observations in $ZY$, LSST $zy$ observations are not critical for all science presented above.  However, some science cases such as LSST photo-$z$ training may require significantly deeper images than the WFD $zy$ observations will produce.

The {\it WFIRST} SN survey is expected to occur in the middle 2 years of the mission.  If {\it WFIRST} is launched in early 2025, the SN survey should occur roughly from 2027--2028.  Our fields (see below) will not be observable with LSST during this entire period.  Nonetheless, we request observations whenever feasible, including possibly reducing the number of filters observed to increase the time baseline (or possibly observing different filters over several consecutive nights).

To prepare for the {\it WFIRST} SN survey, for HLS calibration, and as a jump-start on extragalactic science enabled by the SN survey observations, we expect {\it WFIRST} will obtain relatively deep imaging of the SN fields during commissioning or early in science observations.  We also expect occasional visits to these fields as part of normal survey operations, to monitor variable sources, and as calibration touchstone fields.  We therefore request regular monitoring of the fields, at least twice a year at deep-drilling depths for all years when the {\it WFIRST} SN survey is not running.

Given the current level of uncertainty for {\it WFIRST} launch and
performance, the plan for observing its deep fields should be
reconsidered as the {\it WFIRST} plans develop.

The {\it WFIRST} Project will work to define fields to best fit within
the circular LSST field of view (if {\it WFIRST} SN fields are
$<$10~deg$^{2}$) or multiples of the LSST field of view (if {\it
  WFIRST} SN fields are $>$10~deg$^{2}$).

\vspace{.3in}

\subsection{Footprint -- pointings, regions and/or constraints}

The {\it WFIRST} SN fields must be located in the continuous viewing zone (CVZ) to avoid significant gaps and edge effects in supernova light curves.  The current field of regard is 54$^{\circ}$, constraining the CVZ to declinations of $\delta > |30|$ (the exact constraints are shown in Figure~\ref{f:fields}).  No currently selected deep-drilling field is in the {\it WFIRST} CVZ.  Additional constraints are low Milky Way reddening, low zodiacal light (which is a non-factor in the CVZ), and avoiding bright stars.  These constraints limit the possible overlap with LSST to roughly R.A. between 22 and 6 hours and declinations between $-30$ and $-75$.

While the exact {\it WFIRST} SN field sizes and field centers have not yet been determined, we have selected 4 rough positions, 3 of which are in the LSST footprint.  These fields are at (approximately) $(\alpha, \delta) = (01$:00, -55:00), (04:00, $-35$:00), and (04:44, $-53$:20), with the last field coinciding with the AKARI Deep Field South \citep{Clements12}.  Notably, Euclid has decided to make the AKARI Deep Field South one of their deep fields, thus obtaining deep, high-resolution $RI\!Z$ imaging to 27.2~mag, spectral data with their 0.9--1.2~$\mu$m grism, and long-term (although shallow) light curves in $Y\!J\!H$.

In addition to the above criteria, these regions have significant advantages over other fields.  In particular, they are roughly as far separated as possible within the acceptable southern CVZ, reducing cosmic variance concerns, allowing for isotropy tests, and improving observability from ground-based telescopes.

If the {\it WFIRST} field of regard improves by $\sim$5$^{\circ}$, the Chandra Deep Field South (CDF-S) will be in the CVZ.  Being an existing deep-drilling field and Euclid deep field, having extensive existing multi-wavelength data, and being accessible to Northern-hemisphere telescopes, it is superior to the choices for {\it WFIRST} SN fields listed above.  In the case that the field of regard improves sufficiently, we will likely move the (04:00, $-35$:00) field to CDF-S.  Similarly, if the field of regard increases slightly beyond that, we may shift the (01:00, $-$55:00) field to the existing deep-drilling field centered on the ELAIS deep field.

While the fields have not been precisely defined now, the field of regard will be better defined once the Project enters Phase C (within roughly 1 year).  Furthermore, the {\it WFIRST} SN survey will occur 2 years into the mission, providing sufficient time to determine the in-orbit field of regard before LSST observations are required.  We encourage LSST to continually contact the {\it WFIRST} Project for updates.

\subsection{Image quality}
We have no constraints on image quality.  Cadence is more important.

\subsection{Individual image depth and/or sky brightness}
We request observations that follow the deep-drilling strategy, reaching $u \approx 23.5$, $g \approx 25.3$, $r \approx 25.6$, $i \approx 25.4$, $z \approx 24.9$, and $y \approx 24.0$~mag (depending on the lunar phase).

\subsection{Co-added image depth and/or total number of visits}
We request a cadence matched to the {\it WFIRST} cadence while the fields are available and during the SN survey.  We also request 2 observations per year for the years when the SN survey is not active.

The stacked depth of these fields would be slightly shallower than the final stacked depths of deep drilling fields ($ugrizy = 26.8$, 28.4, 28.5, 28.3, 28.0, 26.2~mag, respectively).  These depths are comparable to the ``shallow'' {\it WFIRST} SN fields (reaching $\sim$28~mag).

\subsection{Number of visits within a night}
We do not require more than 1 visit per night.

\subsection{Distribution of visits over time}
We request a cadence that matches the {\it WFIRST} SN survey (currently planned to be 5 days, but initial attempts to optimize suggest perhaps a 7-day cadence being optimal).  The observations should occur during the {\it WFIRST} SN survey.

We currently make no recommendations for weather loss.  The {\it WFIRST} light curves should be sufficient to constrain the time of peak brightness for SNe.  However, if one wants to use the LSST data independently from {\it WFIRST}, additional observations surrounding (expected) bad weather may be desired.

All filters could be observed in a single visit, but observing all filters over a few nights would be acceptable as long as at least two filters (one color) is observed each night of observations.

\subsection{Filter choice}
We request observations in $ugrizy$.  However, $ugri$ are particularly unique and most important.

\subsection{Exposure constraints}
%\begin{footnotesize}
%{\it Describe any constraints on the minimum or maximum exposure time per visit required (or alternatively, saturation limits).
%Please comment on any constraints on the number of exposures in a visit.}
%\end{footnotesize}
We request deep-drilling style observations.

%\subsection{Other constraints}
%\begin{footnotesize}
%{\it Any other constraints.}
%\end{footnotesize}

\subsection{Estimated time requirement}
%\begin{footnotesize}
%{\it Approximate total time requested for these observations, using the guidelines available at \url{https://github.com/lsst-pst/survey_strategy_wp}.}
%\end{footnotesize}
Scaling from estimates for the deep-drilling fields, but with our
cadence and season duration, we expect each field to require 158
hours, for a total of 474 hours (roughly 1.5\% of total survey time),
if full $ugrizy$ observations are performed.  If observations are
restricted to $ugri$, each field would require 95 hours with a total
of 286 hours (roughly 0.9\% of total survey time) for all fields.

\vspace{.3in}

\begin{table}[ht]
    \centering
    \begin{tabular}{l|l|l|l}
        \toprule
        Properties & Importance \hspace{.3in} \\
        \midrule
        Image quality & 3    \\
        Sky brightness & 2 \\
        Individual image depth & 1  \\
        Co-added image depth & 2  \\
        Number of exposures in a visit   & 2  \\
        Number of visits (in a night)  & 3  \\ 
        Total number of visits & 1  \\
        Time between visits (in a night) & 3 \\
        Time between visits (between nights)  & 1  \\
        Long-term gaps between visits & 1 \\
        Other (please add other constraints as needed) & \\
        \bottomrule
    \end{tabular}
    \caption{{\bf Constraint Rankings:} These rankings primarily reflect our desire to have simultaneous observations with the {\it WFIRST} SN survey.  Cadence/ season duration (and thus number of total visits and gaps between visits) along with individual exposure depth are most important.}
    %Summary of the relative importance of various survey strategy constraints. Please rank the importance of each of these considerations, from 1=very important, 2=somewhat important, 3=not important. If a given constraint depends on other parameters in the table, but these other parameters are not important in themselves, please only mark the final constraint as important. For example, individual image depth depends on image quality, sky brightness, and number of exposures in a visit; if your science depends on the individual image depth but not directly on the other parameters, individual image depth would be `1' and the other parameters could be marked as `3', giving us the most flexibility when determining the composition of a visit, for example.
        \label{tab:obs_constraints}
\end{table}

\subsection{Technical trades}
\begin{enumerate}
    \item For most science, we must reach a depth similar to the {\it WFIRST} observations (both single exposures and stacked depth) to maximize science.  Shallower images will result in a smaller redshift range with overlapping SN and galaxy observations.  Reducing area directly affects the number of objects with data from both telescopes, but is independent of redshift/luminosity.  Reducing observing seasons will reduce the overall stacked depth and produce more extreme ``edge effects'' for the SNe where SN light curves will be cut off.  Because of time dilation, this has an outsized effect on higher-$z$ SNe.
    \item Trading fewer visits for deeper exposures will likely result in insufficient light-curve coverage (especially if weather losses are not mitigated).  Trading individual-exposure depth for additional visits will result in a smaller redshift range (see above) and have an unnecessarily high cadence.
    \item Adjusting the individual exposure times to reach a predetermined depth could be useful as long as the typical depth is not significantly shallower than a constant exposure-time strategy (i.e., the modes are the same).
    \item We could potentially adjust the deep-drilling strategy to spread a set of observations out over a few days to minimize filter changes.  However, this would reduce our ability to have simultaneous colors.
\end{enumerate}

%\begin{footnotesize}
%{\it To aid in attempts to combine this proposed survey modification with others, please address the following questions:
%\begin{enumerate}
%    \item What is the effect of a trade-off between your requested survey footprint (area) and requested co-added depth or number of visits?
%    \item If not requesting a specific timing of visits, what is the effect of a trade-off between the uniformity of observations and the frequency of observations in time? e.g. a `rolling cadence' increases the frequency of visits during a short time period at the cost of fewer visits the rest of the time, making the overall sampling less uniform.
%    \item What is the effect of a trade-off on the exposure time and number of visits (e.g. increasing the individual image depth but decreasing the overall number of visits)?
%    \item What is the effect of a trade-off between uniformity in number of visits and co-added depth? Is there any benefit to real-time exposure time optimization to obtain nearly constant single-visit limiting depth?
%    \item Are there any other potential trade-offs to consider when attempting to balance this proposal with others which may have similar but slightly different requests?
%\end{enumerate}}
%\end{footnotesize}

\newpage
\section{Performance Evaluation}
%\begin{footnotesize}
%{\it Please describe how to evaluate the performance of a given survey in achieving your desired
%science goals, ideally as a heuristic tied directly to the observing strategy (e.g. number of visits obtained
%within a window of time with a specified set of filters) with a clear link to the resulting effect on science.
%More complex metrics which more directly evaluate science output (e.g. number of eclipsing binaries successfully
%identified as a result of a given survey) are also encouraged, preferably as a secondary metric.
%If possible, provide threshold values for these metrics at which point your proposed science would be unsuccessful 
%and where it reaches an ideal goal, or explain why this is not possible to quantify. While not necessary, 
%if you have already transformed this into a MAF metric, please add a link to the code (or a PR to 
%\href{https://github.com/lsst-nonproject/sims_maf_contrib}{sims\_maf\_contrib}) in addition to the text description. (Limit: 2 pages).}
%\end{footnotesize}
Primary metrics:
\begin{itemize}
    \item Total overlapping area with {\it WFIRST} SN deep fields.
    \item Number of deep-drilling depth observations of the fields obtained during the {\it WFIRST} SN survey.
    \item Number of deep-drilling depth observations of the fields obtained outside the time of the {\it WFIRST} SN survey.
    \item Median cadence during the {\it WFIRST} SN survey.
    \item Maximum gap between epochs during and outside the {\it WFIRST} SN survey.
    \item Total stacked depth in each filter.
\end{itemize}

\noindent
Secondary metrics:
\begin{itemize}
    \item Number of SNe Ia detected by LSST in {\it WFIRST} SN deep fields during the {\it WFIRST} SN survey.
    \item Number of $0.45 < z < 0.55$ SNe Ia with light curves that result in distance modulus statistical uncertainty of 0.05~mag.
    \item Number of $0.95 < z < 1.05$ SNe Ia with light curves that result in distance modulus uncertainty of 0.05~mag.
    \item Number density of 5-$\sigma$-detected galaxies at a flux level of $10^{-16}$~erg~s$^{-1}$~cm$^{-2}$.
\end{itemize}
\vspace{.6in}

\section{Special Data Processing}
%\begin{footnotesize}
%{\it Describe any data processing requirements beyond the standard LSST Data Management pipelines and how these will be achieved.}
%
We require nightly co-adds of the LSST data and difference imaging of these co-adds.  While not required as the data are obtained, combined processing of LSST and {\it WFIRST} data will be required for some science cases.  To provide the most science and reach the largest number of researchers, we suggest having access to the {\it WFIRST} and LSST data as well as the {\it WFIRST} and LSST data processing software in a common interactive software environment.

%\section{Acknowledgments}
% \begin{footnotesize}
%R.J.F.\ is supported in part by NASA grant NNG17PX03C, NSF grants AST--1518052 and AST--1815935, the Gordon \& Betty Moore Foundation, the Heising-Simons Foundation, and a fellowship from the David and Lucile Packard Foundation.
% \end{footnotesize}

\newpage

\bibliographystyle{aasjournal}
\bibliography{lsst.bib}

\end{document}